\documentclass{emulateapj}

\newcommand{\be}{\begin{equation}}
\newcommand{\ee}{\end{equation}}

\shorttitle{IMBH and binaries in star clusters}
\shortauthors{Trenti et al.}

\begin{document}


\title{Primordial binaries and intermediate mass black holes in globular
clusters}


\author{M. Trenti,\altaffilmark{1} E. Ardi,\altaffilmark{1}
S. Mineshige,\altaffilmark{1} and P. Hut\altaffilmark{2}}
\altaffiltext{1}{Yukawa Institute for Theoretical Physics, Kyoto
University, 606-8502 Kyoto Japan.}
\altaffiltext{2}{Institute for Advanced Study, Einstein Drive, Princeton, NJ 08540.}

\begin{abstract}


We present the first study of the dynamical evolution of a star
cluster that combines a significant population of primordial binaries
with the presence of a central black hole.  We use direct N-body
simulations, with a black hole mass of about one percent of the total
mass of the cluster.

The evolution of the binary population is strongly influenced by the
presence of the black hole, which gives the cluster a large core with
a central density cusp.  Starting from a Plummer profile, we first
encounter a phase, that last approximately $10$ half-mass relaxation
times, in which binaries are disrupted faster compared to analogous
simulations without a black hole.  Subsequently, however, binary
disruption slows down significantly, due to the large core size.

The dynamical interplay between the primordial binaries and the black
hole thus introduces new features with respect to the scenarios
investigated so far, where the influence of the black hole and of the
binaries have been considered separately.  Specifically, the pattern
of binary destruction by an intermediate-mass black hole may leave a
fingerprint that could be detected observationally.

\end{abstract}


\section{Introduction}

Over the last few years some tantalizing, but yet far from conclusive,
evidence has been accumulating in support of the idea that some star
clusters could harbor a central black hole (BH) with a mass of the order
of $10^3 M_{\sun}$ or more.  Detection of such an intermediate
mass black hole (IMBH) has been claimed for $M15$ and
$G1$ \citep{ger03,geb02,geb05}, although accurate dynamical models of
$M15$ and $G1$ can be obtained without a central BH
\citep{bau03a,bau03b}.  Interestingly, the visual appearance of globulars
containing an IMBH is not that of a so-called core-collapsed cluster,
but rather that of a cluster with a still sizable core \citep{bau04c}.

IMBHs present a high theoretical and observational interest as these
could be potential ultra-luminous X-ray sources and even emit
gravitational waves, detectable by the next generation of gravitational
wave detectors, as a result of close interactions with stars. However,
despite this interest and the fact that theoretical studies of BHs in
stellar systems started more than 30 years ago \citep[e.g.,
see][]{pee72,bah76}, detailed direct N-body simulations to study the
dynamics of an idealized model with single stars and a central BH have
been performed only recently (\citealt{bau04a,bau04b}; see also the
studies on the formation of IMBHs by runaway mergers of massive stars by
\citealt{por04}).

One ingredient that complicates N-body simulations of globular clusters
is the presence of primordial binaries.  Often, these are neglected in
large simulations despite the increasing observational evidence that many
stars in a globular cluster have a companion \citep{hut92,alb01,bel02,zha05}.
This frequent neglect is due to the dramatic increase 
in computational resources required in a simulation where the local
dynamical timescale may be many orders of magnitude smaller than the
global relaxation timescale (hard binaries have an orbital period of a
few hours, while the half-mass relaxation time can be up to a few
billion years). The study of the dynamics in the presence of primordial
binaries has been mainly limited to Fokker-Planck or Monte Carlo
approaches \citep{gie00,fre03} and to direct simulations with rather
modest particle numbers, from $N \approx 10^3$ \citep{mcm90,mcm94,heg92}
to recent higher resolution simulations, with $N$ up to $16384$
\citep{heg05}. Some realistic simulations, including primordial binaries
are available \citep{por04b}, but these are limited only to the first
stage of the life of young dense clusters.  In the case of open clusters,
M67 has been modeled in a 36,000-body simulation running for several
Gigayears \citep{hur05}.

The presence of either an IMBH or a significant population of primordial
binaries leads to an early release of abundant energy, inhibiting the
development of a deep core collapse and hence of the onset of
gravothermal oscillations.  An IMBH can generate energy by swallowing or
tightly binding stars deep in its potential well [note also that energy
can be generated through encounters of stars in the density cusp that is
formed around the BH], while primordial binaries can generate energy by
rapidly increasing binary binding energy through three and four body
encounters.  Energy thus generated in the core of the system fuels the
expansion of the half-mass radius, leading to a self-similar expansion
of the entire system \citep{hen65}.

In this work we present the first direct simulations of the evolution
of globular clusters with both a significant fraction of primordial
binaries (10~\%) as well as an IMBH with a mass of $1.4-2.5~\%$ of the
total mass of the system.  We address the following questions.  Under
the combined effect of the BH and of the binaries, what is the
equilibrium size of the cluster core?  How is the binary
population affected by the presence of the central IMBH? Which
physical processes dominate in the core? In principle two
competing effects are possible: the BH may either enhance the
disruption rate of binaries both indirectly, due to the creation of a
density cusp, and directly, by tidal stripping \citep{pfa05}, or it
may reduce the probability of interactions between binaries and singles,
by producing a low stellar density in a relatively large core
\citep{bau04a}. Which process is dominant is of fundamental
importance. These questions are addressed in the next sections.

\section{Numerical simulations: setup}\label{sec:ns}

The simulations presented in this paper have been performed using the
NBODY-6 code \citep{aar03}, that has been modified with the kind help of
Dr. Aarseth to ensure a more efficient and accurate treatment of the
dynamics around the BH (the relative energy error at the end of our
simulations is below $0.5~\%$). We used a total of $N_{tot}=9011$ equal
mass stars (each of mass $m$), arranged in $819$ binaries and $7373$
singles, plus a BH, introduced as a massive star, with mass ($m_{BH}$)
in the range $1.4-2.5~\%$ of the total mass of the system. The system is
isolated and the initial binding energy distribution for the binaries is
in the range $3$ to $400~kT$ (where $3/2\ kT$ is the average kinetic
energy of a star of the system), flat in logarithmic scale as in the
runs by \citet{heg05}, that we use for comparison, as here we employ the
same initial conditions except for the presence of the BH. The
initial mass distribution is that of a Plummer model. To initialize the
simulation in a situation of approximate dynamical equilibrium in the
presence of the BH we generate a Plummer model made of single stars
only, we then scale the velocities of the particles to reach virial
equilibrium and let the system evolve for $5$ half-mass crossing
times. Some stars are then selected at random and binaries are added to
the simulations, with the standard initialization provided by
NBODY-6. The evolution of the system is then followed up to $t \approx
25~t_{rh}(0)$. For $t_{rh}(0)$ we mean $t_{rh}$ computed at $t=0$. Here
$t_{rh}$ is the half-mass relaxation time \citep{spi87}, defined as
$t_{rh}=0.138 N r_h^{3/2}/(\sqrt{M G} \ln{(0.11N)})$, with $r_h$ being
the half-mass radius, $M$ the total mass of the system and $N$ the
number of centers of mass (i.e. $N=8192$ in our case); in NBODY units
$t_{rh}(0) \approx 112$ for our initial conditions.  

\section{Global evolution}\label{sec:sim}

The large scale structure evolution of the cluster is dominated by the
heating related to the presence of the BH. In fact only the
inner regions of the system experience a mild collapse on a timescale
that, depending on the mass of the BH, is of the order of a few
$t_{rh}$. Inside the core radius $r_c$, defined as the density averaged
radius \citep{cas85}, a cusp in the density profile is formed within the
sphere of influence $r_i$ (with $r_i \approx 15~r_c m_{BH}/M$, see
\citealt{bau04a}) of the BH, with a profile proportional to
$\approx 1/r^{1.7}$ and thus similar to the $1/r^{1.75}$ measured by
\citet{bau04a}. For $m_{BH}=0.014M$ the influence radius is
approximately $0.2r_c$.  By definition, the stellar mass within this
radius is comparable to that of the BH, and thus around one
percent of the total mass of the cluster.

In case of $m_{BH}=0.014M$, the core radius $r_c$ is reduced in $\approx
4~t_{rh0}$ from the initial value of $0.4$ to $0.3$ (in NBODY
units). This is to be compared with a value of $\approx 0.1$ reached
without the BH; when a BH (with the same mass) but no primordial
binaries are present $r_c$ goes down to $\approx 0.28$.

After the first mild contraction all Lagrangian radii start to expand
steadily and a self-similar regime sets in, with the half-mass radius
growing in proportion to $\approx t^{2/3}$, in agreement with the
theoretical argument given by \citet{hen65}. The half-mass radius is
thus marginally bigger (by $\approx 10~\%$ at $24~t_{rh0}$) due to the
presence of the BH; without a BH the half-mass expansion starts only
after core collapse, which takes $\approx 10~t_{rh0}$ in that
case. Conversely, if a BH but no binaries are present, the expansion
rate of $r_h$ is reduced by $\approx 20~\%$. A summary of the properties
of our simulations is reported in Table~\ref{tab1}.

\begin{table}
\begin{center}
\caption{Summary of runs with $N=8192$\label{tab1}}
\begin{tabular}{ccccc}
\tableline\tableline
$m_{bh}/M$ & $f$ & $(r_c)_{cc}/r_{c0}$ & $r_c/r_h$ & $r_{hf}/r_{h0}$ \\ 
\tableline
0.014 & 0.1 & 0.75 & 0.29 & 2.46 \\
0.025 & 0.1 & 0.80 & 0.31 & 2.83 \\
0.014 & 0 & 0.70 & 0.27 & 1.92 \\
0 & 0.1 & 0.25 & 0.09 & 2.25 \\
\tableline
\end{tabular}
\tablecomments{In the first column we report the BH mass, in the second
 the fraction $f$ of primordial binaries, in the third the core radius
 at the end of the initial core contraction phase $(r_c)_{cc}$ in units
 of the initial core radius $r_{c0}$, the fourth entry is the core to
 half mass radius ratio during the self-similar expansion of the system,
 while the last entry ($r_{hf}$) is the value of the half mass radius at
 $t=24~t_{rh}(0)$ in units of the initial half mass radius.}
\end{center}
\end{table}

\section{Properties of the binary population}\label{sec:bin}

\begin{figure}
\plotone{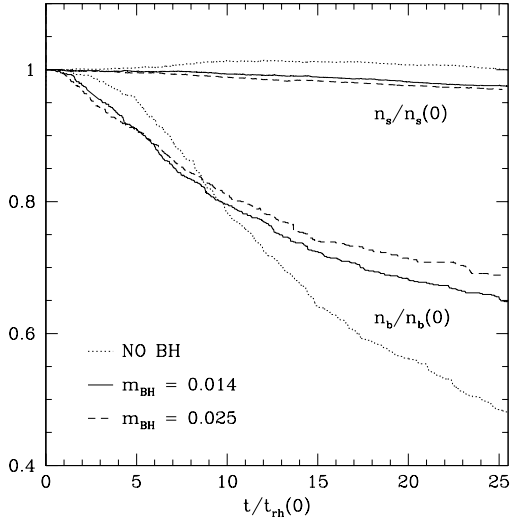} \caption{Time-dependence of the number of
single (upper set of lines) and binary stars (lower set of lines)
expressed as a fraction of the initial values. The data refer to
simulations with $N=8192$ and $10\%$ primordial binaries.  There are
three choices for BH mass: no BH
(dotted line), a central BH with $m_{BH}=0.014M$ (solid line),
and $m_{BH}=0.025M$ (dashed line).  The unit of time is the initial
half-mass relaxation time.\label{fig1}}
\end{figure}

As the presence of the BH dominates the global dynamics of the
cluster, the evolution of the binaries presents some remarkable
differences from the scenario where no central BH is present.
We can distinguish two phases.

The presence of a BH significantly accelerates the rate at which
binaries are disrupted in the first few half-mass relaxation times
(see Fig.~\ref{fig1}).  The reason is that binaries that pass near the
BH can be quickly destroyed. The disruption rate is approximately
constant during the first $5~t_{rh0}$, while without a BH disruption
takes a while to get underway. The initial
disruption rate depends on the mass of the BH: a more
massive BH starts burning binaries at a higher rate, as can be seen from
Fig.~\ref{fig1}.

For our simulation with $m_{BH}=0.014 M$ we observe a binary depletion
rate of $\approx 0.15$ binaries per initial half-mass crossing time,
during the first $5~t_{rh0}$ ($80$ binaries in total).  Around half of
the binaries are disrupted within $r_i$. Interestingly only two of them
happen to be so close to the BH tidal radius $r_t$ that the disruption
may be considered a direct consequence of the tidal stripping force
exherted by the BH. Here $r_t=(m_{BH}/m)^{1/3} a \approx 5 a$, where $a$
is the semi-major axis of the binary.  With this respect, the
theoretical model recently proposed by \citet{pfa05} applied to our
simulation would give (from his Eqs.~11-12) a tidal disruption rate of
$\approx 2 \cdot 10^{-2}$ binaries per initial half-mass crossing time,
so that we observe less tidally stripped binaries than expected
($\approx 10$). The disruption of binaries by the BH is thus mainly due
to an indirect effect: binaries venturing close to the BH, where
the density is higher, are more likely to interact with a single star or
with another binary. In fact in our simulations we often observe the
presence of hierarchical systems within $r_i$, with the BH playing the
role of a perturber.

After the initial transient phase a self-similar expansion sets in,
where the average core density is much less (approximately by a factor
$10$) than it would be without the presence of a central compact object.
In this second phase we observe (see Fig.~\ref{fig1}) a reduced rate of
binary disruption (as this rate is proportional to the square of the
density, e.g. see \citealt{ves94}), so that the difference between a
simulation with and without a BH becomes remarkable. The turning point
is around $10~t_{rh0}$, when the number of surviving binaries for a
simulation with a central object becomes greater than in the absence of
a BH. Our Fig.~\ref{fig1} has been given in units of the initial
half-mass relaxation time, but the picture remains qualitatively the
same even if we consider a co-moving time coordinate to take into
account the differences between the half-mass radii of a simulation with
and without a BH.

\begin{figure}
\plotone{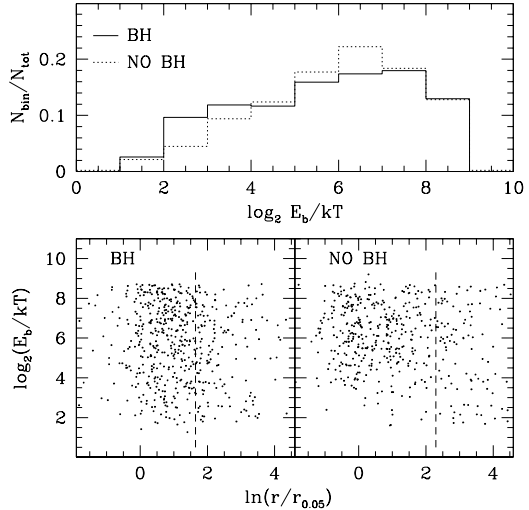} \caption{Distribution of binding energies for
binaries (upper panel) at time $t=24~t_{rh}(0)$ for a simulation with
$N=8192$, $10~\%$ primordial binaries and $m_{BH}=0.014M$ (solid
line), compared to a similar simulation without the BH (dotted
line). In the lower two panels we report the energy-radius
distribution of binaries for the same simulations. The radii have been
normalized to the Lagrangian radius containing $5~\%$ of the mass in
the simulation. The dashed line gives the position of the half-mass
radius, at $t=24~t_{rh0}$. The relatively large survival at
$t=24~t_{rh0}$ of binaries within the half-mass radius with binding
energy $E_b<20kT$ is clearly visible in the bottom panel, and is also
reflected in the higher values for the solid line at the left-hand
side of the upper panel.\label{fig2}}
\end{figure}

Interestingly the spatial distribution of binaries is also different
from the case without a BH: as can be seen from
Fig~\ref{fig2}, the number of binaries within the $0.05$ Lagrangian
radius is much less for the simulation with a BH. This is
probably due to the disruption of binaries that approach close
to the BH. In the presence of a BH, binaries tend to be more
concentrated between the $0.05$ Lagrangian radius and the half-mass
radius, while in the absence of a BH, binaries can sink deeper into the central
region of the cluster. As a result, the ratio of binaries to singles in
the core of the system is between two and three times less than in
absence of the BH.

If we compare the evolution of the binding energy distribution of the
binaries in a run with and without a central BH, we can see that
the net effect of the BH at later times is to somewhat enhance
the survival probability of binaries with binding energies of a few
$kT$s ($E_b \lesssim 16~kT$), especially for binaries within the
half-mass radius (see Fig.~\ref{fig2}). This feature is especially
apparent in a two-dimensional plot of the distribution of binaries as
a function of binding energy and radial distance from the cluster center,
depicted in the bottom panels of
Fig.~\ref{fig2}. The rather strong correlation between radius and
binding energy that is observed \citep[e.g., see][]{gie00,heg05} in
systems without a central compact object is thus almost absent.

\section{Discussion and conclusions}\label{sec:conclusion}

In this paper we have presented for the first time the results of direct
N-body simulations of the evolution of a globular cluster with a
population of primordial binaries that harbors a central IMBH, with a
mass of the order of $1~\%$ of the total mass of the system. In order to
begin to analyze the basic evolutionary processes we have restricted our
simulations to isolated systems of stars of equal mass, while for the
time being neglecting stellar evolution and physical collisions.

The environment around the BH turns out to be of great interest from a
dynamical point of view. It represents a laboratory where complex
interactions between hierarchical systems take place. Around the BH we
observe usually one or more stars tightly bound to it. With a
significant population of primordial binaries, it frequently happens
that a binary approaches the center of the system.  Interactions between
these subsystems are important not only from a dynamical point of view,
but also because they form a factory to produce exotic objects, such as
tight (X-ray emitting) binaries and high velocities escapers, which we
observe in high numbers in our simulations. In our run with
$m_{BH}=0.014 M$ we find that a fraction of $3 \cdot 10^{-3}$ of the
stars of the cluster leave the system in a relaxation time and that
$\approx 1/3$ of them have escape velocities that exceed by at least
five times the core velocity dispersion. Of these $54$ were binaries; in
addition one triple was ejected from the system.

From a computational point of view the complex interactions between
binaries and the BH have proved to be very challenging for current
state of the art N-body algorithms. Clearly, new methods have to be
developed to provide both more accuracy and higher efficiency.

We have shown that a BH with a mass of around $1~\%$ of the cluster
mass can produce most of the energy required to fuel the expansion of
the globular cluster, so that the evolution of the large scale
structure of a globular cluster with primordial binaries and a central
BH is remarkably similar to that attained in the absence of binaries
(see \citealt{bau04a}).  Even so, the evolution of the population of
primordial binaries is strongly influenced by the BH, so that in
principle indirect evidence for the presence of the BH can be obtained
by observing the distribution of binaries.

Observational detection of an unexpectedly large number of only
moderately hard binaries in the core could be used as arguments for
the presence of a IMBH in old star clusters, i.e. in those systems
whose age would imply a depletion in the central region of binaries
with binding energy below $\approx 15~kT$.  These binaries are able to
survive in significant numbers when an IMBH is present.

Clearly, a quantitative comparison with observations would require us
to proceed beyond the simple models presented here. To start with, we
would need to include a realistic number of stars to avoid possible
biases introduced by the scaling of the ratio of the BH mass to that
of single stars. In addition, the introduction of a realistic initial
mass function is likely to modify the concentration of binaries in the
center of the system due to mass segregation, and stellar evolution will
also influence the distribution of binary binding energy.  However,
the main effects presented in this letter are likely to be present, at
least in qualitative ways, in more detailed realistic simulations.

\acknowledgments

We are indebted to Sverre Aarseth for providing his code NBODY-6 with
ad-hoc modifications in order to make possible running the simulations
presented in this work. We thank Holger Baumgardt for interesting
discussions. This work is supported in part by the Grants-in-Aid of the
Ministry of Education, Science, Culture, and Sport, (14079205; MT, EA,
SM) and by a Grant-in-Aid for the 21st Century COE {\lq\lq}Center for
Diversity and Universality in Physics" (SM). P.H. thanks Prof. Ninomiya
for his kind hospitality at the Yukawa Institute at Kyoto University,
through the Grants-in-Aid for Scientific Research on Priority Areas,
number 763, "Dynamics of Strings and Fields", from the Ministry of
Education, Culture, Sports, Science and Technology, Japan. The numerical
simulations have been performed on the Condor cluster at the Institute
for Advanced Study.

\end{document}